\documentclass[ejp]{iopart}
\usepackage{graphicx}
\begin{document}
\title{Another look through Heisenberg's microscope}

%
%

\author{Stephen Boughn$^{1,2}$ and Marcel Reginatto$^{3}$}

\address{$^{1}$Department of Physics, Princeton University, Princeton, NJ,\\ $^{2}$Departments of Physics and Astronomy, Haverford College, Haverford, PA,\\$^{3}$ Physikalisch-Technische Bundesanstalt, Braunschweig, Germany}

\ead{sboughn@haverford.edu}


\begin{abstract}
Heisenberg introduced his famous uncertainty relations in a seminal 1927 paper entitled {\it The Physical Content of Quantum Kinematics and Mechanics}. He motivated his arguments with a {\it gedanken} experiment, a gamma ray microscope to measure the position of a particle.  A primary result was that, due to the quantum nature of light, there is an inherent uncertainty in the determinations of the particle's position and momentum dictated by an indeterminacy relation, $\delta q \delta p \sim h$.  Heisenberg offered this demonstration as ``a direct physical interpretation of the [quantum mechanical] equation $\textbf{pq} - \textbf{qp} = -i\hbar$''
but considered the indeterminacy relation to be much more than this. He also argued that it implies limitations on the very meanings of position and momentum and emphasized that these limitations are the source of the statistical character of quantum mechanics.  In addition, Heisenberg hoped but was unable to demonstrate that the laws of quantum mechanics could be derived directly from the uncertainty relation.
In this paper, we revisit Heisenberg's microscope and argue that the Schr\"{o}dinger equation for a free particle does indeed follow from the indeterminacy relation together with reasonable statistical assumptions.
\end{abstract}


\section{Introduction}

The idea of fundamental uncertainty in nature was introduced by Werner Heisenberg in his seminal 1927 paper entitled {\it \"{U}ber den anschaulichen Inhalt der quantentheoretischen Kinematik und Mechanik}, translated by Wheeler and Zurek as ``The Physical Content of Quantum Kinematics and Mechanics'' \cite{H27}. The German noun {\it Anschaulichkeit} and the adjective {\it anschaulich} connote visualization or intuition through mechanical models \cite{M84}. While Wheeler and Zurek translate the German words ``{\it anschaulichen inhalt}'' as ``physical content,'' the historian and philosopher of science Miller favors ``intuitive contents''
\cite{M84,M94}.
Heisenberg's biographer Cassidy prefers the  expression ``perceptual content'' instead \cite{C92}.  Thus there are various alternative ways of translating the title of Heisenberg's paper. {\it Anschaulich} seems to be ``one of those German words that defy an unambiguous translation'' as Uffink has pointed out \cite{U02}.

Heisenberg's use of the word {\it anschaulich} was undoubtedly intended to answer Schr\"{o}dinger's argument that his wave mechanics was more {\it anschaulich} than Heisenberg's matrix mechanics \cite{HU14}.  However, Heisenberg's paper turned out to be much more than a defense of the matrix mechanics formulation of quantum mechanics.  The very first sentence in his paper declares, ``We believe we understand the physical content of a theory when we can see its qualitative experimental consequences in all simple cases and when at the same time we have checked that the application of the theory never contains inner contradictions'' \cite{H27}.  He then proceeds, largely through the use of a simple gedanken experiment, a $\gamma$-ray microscope, to give a simple physical interpretation of the abstract quantum mechanical commutation relation, $\textbf{pq} - \textbf{qp} = -i\hbar$.  By the end of the paper, he was able to conclude that ``as we can think through qualitatively the experimental consequences of the theory in all simple cases, we will no longer have to look at quantum mechanics as unphysical and abstract'' \cite{H27}.  This was an extraordinary claim about a theory that nearly all physicists, at the time, characterized as unphysical and abstract.

In fact, soon after Heisenberg's paper was received, there arose many questions of clarification and outright objections to his conclusions.  Most notably, Bohr criticized several aspects of Heisenberg's treatment even before the paper was published.  Curiously, Heisenberg included these criticisms at the end of his paper in an {\it Addition in Proof} but did not address them in the body of the paper.  While Bohr acknowledged the importance of Heisenberg's indeterminacy relations, he considered them to be an aspect of his principle of complementarity, a principle Bohr considered to be at the heart of quantum mechanics.  In particular, he pointed out that Heisenberg's analysis of the $\gamma$-ray microscope presupposed complementarity (in this case, the wave/particle duality of photons).  He also pointed out a flaw in Heisenberg's analysis of the optics that, when corrected, did not change the conclusions. The importance of complementarity was later stressed by Bohr \cite{B28}, Heisenberg \cite{H30} and Pauli \cite{P33}. This way of looking at the Heisenberg relations became the standard view of the Copenhagen interpretation.

The modern consensus is that Heisenberg's derivation is, at best, a heuristic argument and the uncertainties that appear in the standard uncertainty principle $\Delta q \Delta p \geq \hbar/2$, which follows immediately from the commutation relation for $\textbf{q}$ and $\textbf{p}$, have little to do with disturbance and measurement errors \cite{B70,B98,BR80}.
Rather $\Delta q \Delta p \geq \hbar/2$ simply expresses a property of all quantum states.  On the other hand, even this interpretation cannot be divorced from the language of experiment.  Systems must be prepared in quantum states and {\it state preparation} necessarily involves the action of some experimental apparatus. Furthermore the meaning of, for example, $\Delta q$ is the spread in the results of the {\it measurements} of $q$ for similarly prepared systems and this also involves the action of an experimental apparatus. However, the uncertainty principle does not necessarily restrict the {\it accuracy} with which $q$ and $p$ can be simultaneously measured; in fact, it can be argued that quantum mechanics says nothing at all about the simultaneous measurements of non-commuting observables 
\cite{B70}

Nevertheless, the question of how to relate Heisenberg's analysis to the formalism of quantum mechanics has never been completely settled and discussions about the meaning of and primacy of Heisenberg's uncertainty relations have continued unabated.  These discussions often focus on one or more of the different expressions of the uncertainty principle such as: restrictions on the accuracy of simultaneous measurements of canonically conjugate quantities, e.g, $p$ and $q$; restrictions on the of the spread of individual measurements of conjugate quantities made on an ensemble of similarly prepared systems; restrictions on the physical compatibility of experimental arrangements for accurately measuring different observables; and the inevitable disturbance of a system due to its interaction with a measuring device.  We will comment on some of  these  in Section 2. Heisenberg did not seem to distinguish among the above types of uncertainty in his paper nor did he offer a quantitative definition of the uncertainties $\delta q$ and $\delta p$ themselves.   It seems to us that this was entirely in line with his intention of providing an {\it anschaulich} description of quantum mechanics with the purpose of providing physical insight into the content of the abstract formalism.  This certainly seems to be the purpose of many authors of introductory QM texts who include descriptions of Heisenberg's $\gamma$-ray microscope.

The purpose of our present paper is to push Heisenberg's argument further and use his $\gamma$-ray microscope not just to provide ``a direct physical interpretation'' of the quantum mechanical formalism but to show that his qualitative indeterminacy relation, $\delta q \delta p \sim h$,  complemented with assumptions about the statistical analysis of measurement, provides a direct route to the free particle Schr\"{o}dinger equation (see Section 3).  Heisenberg considered the possibility of such a derivation in the conclusions of his paper where he wrote ``Of course we would also like to be able to derive, if possible, the quantitative laws of quantum mechanics directly from the physical foundations - that is, essentially, from relation (1) [$\delta q \delta p \sim h$]'' but then was forced to conclude that ``We believe, rather, for the time being that the quantitative laws can be derived out of the physical foundations only by use of the principle of maximum simplicity'' \cite{H27}.
Near the end of his paper, Heisenberg states
\begin{quote}
We have not assumed that quantum theory -- in opposition to classical
theory -- is an essentially statistical theory in the sense that only statistical
conclusions can be drawn from precise initial data... Even in principle
we cannot know the present in all detail.  For that reason everything
observed is a selection from a plenitude of possibilities and a limitation
on what is possible in the future.
\end{quote}
In his 1929 Chicago lectures, ``The Physical Principles of the Quantum Theory''
\cite{H30},
he pointed out that
\begin{quote}
...the idea that natural phenomena obey exact laws -- the principle of causality (...) rests on the assumption that it is possible to observe the phenomena without appreciably influencing them.
\end{quote}
However, nature is quantized and
\begin{quote}
There exists no infinitesimals by the aid of which an observation might be made without appreciable perturbation.
\end{quote}

\section{Heisenberg's Microscope and the Indeterminacy Relations}

Near the beginning of his paper, Heisenberg made use of a $\gamma$-ray microscope to investigate the meaning of position and momentum of a particle in the context of quantum mechanics.  Since Einstein's special theory of relativity, physicists have learned to pay strict attention to the degree to which theoretical quantities can be measured.  Bohr emphasized this in his debates with Einstein and Heisenberg certainly had this in mind when he wrote \cite{H27}
\begin{quote}
When one wants to be clear by what is to be understood by the words ``position
of the object,'' for example of the electron (relative to a given frame of reference),
then one must specify definite experiments with whose help one plans to measure
the ``position of the electron''; otherwise, this word has no meaning.
\end{quote}
Heisenberg then proceeds to introduce a $\gamma$-ray microscope with which one can determine the position of an electron with arbitrary accuracy so long as the wavelength of the $\gamma$-ray is small enough.  In general, one can determine the position to an accuracy on the order of the wavelength $\lambda$ of the $\gamma$-ray, i.e., $\delta q \sim \lambda$.  However, in making this measurement, the scattering of the $\gamma$-ray by the electron imparts a momentum impulse to the electron proportional  to the momentum of the $\gamma$-ray photon, which by the Einstein relation is $h/\lambda$.  If one considers this impulse as an unknown momentum disturbance, then the uncertainty in the electron's momentum is $\delta p \sim h/\lambda$ (assuming the initial momentum of the electron is known precisely).  Combining these two relations one has											 
\begin{equation}
\delta q \delta p \sim h
\end{equation}
Later in the paper he made the analogy with relativity explicit \cite{H27}:
\begin{quote}
It is natural in this respect to compare quantum theory with special relativity.  According to relativity, the word ``simultaneous" cannot be defined except through {\it experiments} in which the velocity of light enters in an essential way.  If there existed a ``sharper" definition of simultaneity, as, for example, signals which propagate infinitely fast, then relativity theory would be impossible...We find a similar situation with the definition of the concepts of ``position of an electron" and ``velocity" in quantum theory.  All {\it experiments} which we can use for the definition of these terms necessarily contain the uncertainty implied by equation (1) [$\delta q \delta p \sim h$]...
\end{quote}
So, it seems, that Heisenberg considered the indeterminacy relation to be the result of an empirical principle embodied in his $\gamma$-ray microscope in the same way that the relativity of simultaneity was due to the empirical principle that the speed of light was a universal constant.  For Heisenberg Eq. 1, the first and most important relation in his paper, imposes fundamental limitations on the very meaning of position and momentum.  He concludes that ``...Even in principle we cannot know the present [state of a system] in all detail.''

Heisenberg's paper ends with a remarkable {\it Addition in Proof} in which he points to a number of criticisms by Bohr. As Rosenfeld would later argue in his article on the history of atomic theory \cite{R71}, Heisenberg
\begin{quote}
declared in substance that he had missed essential points, whose clarification would be found in a forthcoming paper by Bohr. This addendum must have puzzled many readers: it is not often that the announcement of a decisive progress in our insight into the workings of nature is qualified by such a warning.
\end{quote}
Bohr, in a subsequent paper in {\it Nature} \cite{B28}, would emphasize the primacy of the Principle of Complementarity in interpreting Heisenberg's uncertainty relations. A year later Heisenberg, in his 1929 Chicago lectures \cite{H30}, corrected some technical errors (which however did not affect the conclusions of his 1927 paper) and acknowledged the importance of the complementary nature of wave and particle descriptions. However, as Camilleri has emphasized, Heisenbergs's understanding of complementarity differed in some important respects from Bohr's view \cite{C09}. The extent to which complementarity sets limits on conjugate quantities has been clarified by Hall \cite{H04}, who succeeded in deriving generally applicable uncertainty relations which quantify the limitations imposed by complementarity on quantum systems.

While there has never been complete agreement on just what the standard Copenhagen interpretation of quantum mechanics entails \cite{S72,Ho04,C09}, Pauli's 1958 article, ``General Principles of Quantum Mechanics'' \cite{P33} in {\it Handbuch der Physik} might be taken as the ``final'' Copenhagen view on the uncertainty principle. In his discussion, Pauli reproduced both the derivation of Bohr (assuming wave/particle duality of photons) and the (corrected) derivation of Heisenberg. (Pauli also considered the relativistic regime.) The uncertainty relation, $\delta q \delta p \sim h$, for photons follows directly from their wave nature. Photon-particle scattering experiments then transfer this same relation to the position and momentum of particles. To Pauli, the simplest interpretation of the latter is that particles also possess wave-like properties such that $\lambda = h / p$. 
Ultimately, Pauli seems to give comparable importance to the uncertainty principle and complementarity, e.g.,
\begin{quote}
The influence of the apparatus for measuring the momentum (position) of the system is such that within the limits given by the uncertainty relationships the possibility of using a knowledge of the earlier position (momentum) for the prediction of the results of the later measurements of the position (momentum) is lost. If, due to this, the use of a classical concept excludes that of another, we call both concepts (e.g., position and momentum co-ordinates of a particle)
{\it complementary} (to each other), following Bohr. We might call modern quantum theory as ``The Theory of Complementarity'' (in analogy with the terminology ``Theory of Relativity'').
\end{quote}

A corrected simple model  of Heisenberg's microscope, which is the one that invariably appears in introductory texts, is depicted in Fig. 1.  For simplicity assume that the $\gamma$-ray photon beam is initially along the axis of the lens.  The angular resolution of a lens of diameter D is $\delta \theta \approx \lambda/D$ (or 1.22 $\lambda/D$ according to the Rayleigh criterion).  This corresponds to a distance at the object of $\delta q \approx F \delta \theta \approx \lambda F/D$ where $F$ is the focal length of the lens.  The photon that is (elastically) scattered into the lens acquires a transverse component that is in the range of $-p_\gamma \sin\phi$ to $+p_\gamma \sin\phi$ where $p_\gamma$ is the original momentum of the photon. For small angles, $\sin\phi \approx \tan\phi=D/2F$. Therefore, the uncertainty in the transverse component of the momentum of the recoiling photon is $p_\gamma \approx (D/F) p_\gamma \approx hD/\lambda F$.  By conservation of momentum, the uncertainty of the electron momentum becomes $ \approx hD/ \lambda F$.  Multiplying the expressions for $\delta q$ and $\delta p$ we obtain $\delta q \delta p \sim h$, the same expression as in Eq. 1.  While this analysis would presumably have passed muster with Bohr, we suspect that Heisenberg was satisfied with his original argument as he maintained that ``…one does not need to complain that the basic equation (1) contains only qualitative predictions.''   He evidently thought such complaints were irrelevant to his argument.

\begin{figure}
  \centering
  \includegraphics[width=5.5cm]{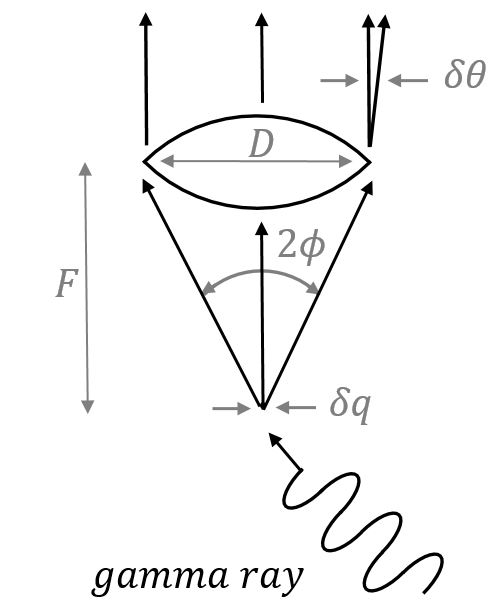}
  \caption{Heisenberg microscope}
  \label{fig1}
\end{figure}

One of the main goals of Heisenberg's paper was to obtain ``a direct physical interpretation of the equation $\textbf{pq} - \textbf{qp} = -i\hbar$.''  To do this he constructed the equivalent of a minimum uncertainty Gaussian wave packet for which the quantum mechanical operators $\textbf{p}$ and $\textbf{q}$ have their usual meanings.  His conclusion was that for this particular quantum state, $\delta q \delta p = \hbar/2 $  if $\delta q$ and $\delta p$ are interpreted as rms expectation values.  Considering the qualitative nature of Eq. 1, Heisenberg then took it to be a direct consequence of $\textbf{pq} - \textbf{qp} = -i\hbar$, the commutation rule of quantum formalism.

Just four months after Heisenberg's paper, Kennard \cite{K27} proved that for any normalized state,
\begin{equation}
\Delta q \Delta p \geq \hbar/2
\end{equation}
where $\Delta q$ and $\Delta p$ are standard deviations, i.e.,  $(\Delta q)^2 \equiv \int \Psi^* q^2 \Psi dq - \left[ \int \Psi^* q \Psi dq \right]^2$ and similarly  for $(\Delta p)^2$. This is the familiar standard expression for the uncertainty principle.  Robertson \cite{R29} and Schr\"{o}dinger \cite{S30} generalized this relation for any pair of observables but their more general expression reduces to Eq. 2 for conjugate pairs \cite{B70}.  We note that in Heisenberg's 1927 paper, his indeterminacy relation is expressed as either an approximate equality or an equality rather than an inequality as in Eq. 2.

There have been many, many analyses of the uncertainty relation since 1927. Most of them take standard, usually nonrelativistic, quantum mechanics  as their starting point and go on to compute  various expressions of  uncertainties that may be derived  from the theoretical formalism.  These analyses  can provide deep insight into the quantum formalism and its interpretation but for the most part they are not directly relevant to the present paper, nor to Heisenberg's original paper for that matter.  It was Heisenberg's intent to derive a relation from an {\it empirical} principle, in this case one governing the behavior of a $\gamma$-ray microscope, and then to compare this relation with one derived from abstract quantum formalism thereby rendering a degree of {\it anschaulichkeit} to quantum mechanics.

The properties of light and electrons that are used for Heisenberg's analysis all derive from experimental results. This supports the view that the indeterminacy relation should be considered an empirical principle, even if  it is unlikely that  Heisenberg's analysis would have been carried out if not for the urgent need to clarify the interpretation of quantum mechanics.  The wave-like behavior of light had been known since Thomas Young's experiments in the 1790's.  The quantal nature of light was known from the work of Planck and Poincare, the analysis of the photoelectric effect by Einstein, and the scattering experiments of Compton.  That electromagnetic radiation causes pressure and therefore has momentum was known  to  Maxwell and others well before special relativity \cite{M91}.

In the last few years, there have been conflicting claims regarding the question of whether Heisenberg's indeterminacy relation $\delta q \delta p \sim h$ is generally applicable; for discussions, see e.g. Refs.
\cite{HU14,S14,R15}.
The validity of the uncertainty principle $\Delta q \Delta p \geq \hbar/2$, which follows from the commutator of $\textbf{q}$ and $\textbf{p}$, is of course not questioned.  But the validity of the relation $\delta q \delta p \sim h$ depends on how one defines uncertainty and disturbance. The papers addressing this controversy consider the problem of taking the rather vague concepts of uncertainty and disturbance that Heisenberg used in his gedanken experiment and replacing them by definitions that are precise and at the same time general enough to apply to a wide class of measurements. Thus inequivalent measurement-disturbance relations can be found in the literature \cite{O03,BLW13}, as discussed in Refs. \cite{HU14,S14,R15} and, not surprisingly, these different definitions can lead to conclusions that are in disagreement.  This is at the moment an active area of research and it is still not clear which of the different Heisenberg-type inequalities that follow from these different definitions will prove useful in the analysis of experiments or for the development of new theoretical tools. In all such studies, an effort is made to find reasonable mathematical expressions derived from the quantum formalism that may be interpreted as measures of uncertainty and disturbance and to look for inequalities which are satisfied by these expressions. Thus the approach of these papers is in some sense a reversal of the procedure followed by Heisenberg, who arrived at his indeterminacy relation after suggesting plausible but imprecise definitions of uncertainty and disturbance for particular experiments.  It is for this reason that the conclusions of these studies are not directly relevant to our paper, which takes Heisenberg's heuristic approach as its starting point to derive additional results. We are not as much interested in what standard quantum formalism has to say about the uncertainty principle but rather in how Heisenberg's empirical principle points the way to quantum mechanics.  It is to this topic that we now turn.

\section{Schr\"{o}dinger equation from the Heisenberg relations}

Einstein, in his ``Reply to Criticisms'' in the volume ``Albert Einstein: Philosopher-Scientist'' \cite{S49}, makes reference to Heisenberg's uncertainty relation in two places. In the first comment, he writes that the correctness of the uncertainty relation ``is, from my own point of view, rightfully regarded as finally demonstrated.'' Later on, he refers to the ``natural limits fixed by the indeterminacy relation'' while discussing ``the very important progress which the statistical quantum theory has brought to theoretical physics.'' These two remarks on the uncertainty relation suggest that he believed it should be taken as an empirical principle, especially given the fact that much of what he writes about quantum mechanics in this essay consists of incisive criticism of the standard interpretation of the theory.

As we mentioned above, Heisenberg had hoped his uncertainty relation would serve as a foundation for quantum mechanics.  It would have been possible, already in the late 1920s, to fulfill his expectations; i.e., to show that the basic equations of quantum mechanics follow from the empirical relation $\delta q \delta p \sim h$.  Such an argument requires only the realization that the standard Hamilton-Jacobi formulation of classical mechanics can serve as the basis for a theory of the motion of classical ensembles \cite{L52,HR16}
and a familiarity with the foundations of statistics. 
These tools are needed to define the uncertainties for position and momentum, which remain vague in Heisenberg's paper, in a more precise way.

Hall and Reginatto \cite{HR02a,HR02b} have already shown that it is possible to go from the Heisenberg uncertainty relation to the Schr\"{o}dinger equation provided one introduces a more powerful {\it exact} form of the uncertainty principle which postulates that the quantum system is subject to momentum fluctuations with a strength that is inversely correlated with the uncertainty in position. They argued that their approach offered \cite{HR02a}
\begin{quote}
 ...a new way of viewing the uncertainty principle as {\it the} key concept in quantum mechanics. While it is true that no one before quantum mechanics would think of taking an uncertainty principle as a fundamental principle, our analysis is valuable in that it enforces the importance of the uncertainty principle in distinguishing quantum mechanics from classical mechanics -- in a sense, it says that the uncertainty principle is {\it the} fundamental element that is needed for the transition to quantum mechanics.
\end{quote}

While the approach presented in this paper is in some ways similar to that of Hall and Reginatto \cite{HR02a,HR02b}, it is less formal and perhaps more empirically motivated in the sense that it takes as its starting point the particular type of measurement considered by Heisenberg. To carry out our approach, we will first introduce uncertainties for position and momentum using an operational point of view. Then a physically motivated jump to the Schr\"{o}dinger equation will be made. This requires introducing an additional assumption that effectively amounts to postulating a quantization procedure.

\subsection{The uncertainty in position}

In his discussion of uncertainty in position, Heisenberg does not define precisely what he means. He simply writes (using our notation) ``Let [$\delta q$] be the precision with which the value [$q$] is known ([$\delta q$] is, say, the mean error of [q])'' and then proceeds with the analysis. Here instead we pay close attention to some technical aspects of the measurement of position and derive a precise definition of $\delta q$.

Consider an experiment in which the measurand (i.e., the quantity to be measured) is the position of the particle. We introduce a model relating the value of the measurand $\tilde{q}$ to the estimated value $q_k$ that results from a given measurement by setting
\begin{equation}
q_k = \tilde{q} + \epsilon_k ,
\end{equation}
where the subscript $k=1,...,N$ labels a particular measurement out of a set of $N$ measurements (that is, we allow for the possibility of repeating the measurement $N$ times) and $\epsilon_k$ is the discrepancy between the value of the measurand $\tilde{q}$ and the estimated value $q_k$. Thus the result of the experiment will be a series of numbers, $(q_1,q_2,...,q_N)$, and from this series of numbers the experimentalist will estimate the position of the particle. This requires a mathematical model of the  analysis of the  measurement data. More formally, we assume that there is some probability density $P$ which describes the probability of measuring $q_k$ and treat the measurement $q_k$ as if it was a sample from this distribution. Since $q_k$ is a location parameter, it will be natural to assume that the probability density is of the form $P=P(q-\tilde{q},\lambda_a)$, where the $\lambda_a$, $a=1,...,n$, are any additional parameters which are needed to fully describe the probability density. For example, $P$ might be the Gaussian
\begin{equation}
P = \frac{1}{\sqrt{2 \pi}\sigma}e^{-\frac{1}{2}\left(\frac{q-\tilde{q}}{\sigma}\right)^2},
\end{equation}
in which case there is only one additional parameter $\lambda_1$, which corresponds to the standard deviation, $\lambda_1 = \sigma$. One may, of course, consider more complicated densities which depend on multiple parameters.

The particle's position can be estimated from the data by means of an ``estimator'' or ``point estimate,'' which is a statistic (that is, a function $f(q_k)$ of the data $\{q_k\}$) used to infer the value of the measurand $\tilde{q}$ that enters into the statistical model. There are of course many different possible estimators, some better, some worse. For the analysis of the $\gamma$-ray microscope, let us assume that the following two conditions are required:
\begin{enumerate}
\item As the Heisenberg uncertainty relation is assumed to hold only for the case of an optimal experiment (i.e., a cleverly designed experiment and the best available data analysis), the estimator used should be the one that has the lowest possible variance,
\item the estimator should be an unbiased estimator, that is, the mean of the sampling distribution of the statistic should be equal to the parameter being estimated.
\end{enumerate}
While the second condition seems natural, one should be aware that there is an element of arbitrariness in the criterion of being unbiased
\cite{Silvey,Jaynes,vdLFvT14}.
It is introduced here for technical reasons: it is possible to establish a lower bound for the variance of an unbiased estimator but there is no analogous proof for the general case
\cite{Silvey,vdLFvT14}.
Furthermore, one can argue that this condition is not unreasonable in the context of a repeatable experiment which may be carried out a large number of times.

These assumptions are all that we need for a precise definition of $\delta q$. It can be shown that all unbiased estimators satisfy the Cram\'{e}r-Rao inequality \cite{Bulmer},
\begin{equation}\label{cr}
        var(f) \geq \left\{ \int dq \, P \left( \frac{1}{P}  \frac{\partial P}{\partial \tilde{q}} \right)^2 \right\}^{-1}.
\end{equation}
where $var(f)$ denotes the variance of the estimator $f$. The proof requires a mild regularity condition which allows reversing the order of integration with respect to $q$ and differentiation 
with respect to $\tilde{q}$ \cite{Bulmer} and we assume here that this condition holds. Because $P=P(q-\tilde{q},\lambda)$, we have $\left( \frac{\partial P}{\partial \tilde{q}} \right)^2=\left( \frac{\partial P}{\partial q} \right)^2$ and the inequality can be written in the equivalent form
\begin{equation}\label{cr1}
        var(f) ~ \geq \left\{ \int dq \, P \left( \frac{1}{P}  \frac{\partial P}{\partial q} \right)^2 \right\}^{-1}.
\end{equation}

Assume that $\tilde{q}$  is estimated from the data using the {\it best possible estimator}, which we take to mean the unbiased estimator that has the lowest variance, and that this estimator achieves the Cram\'{e}r-Rao lower bound (i.e., it is an `efficient estimator'). Then,
\begin{equation}\label{deltax}
        (\delta q)^2 = \left\{ \int dq \, P \left( \frac{1}{P}  \frac{\partial P}{\partial q} \right)^2 \right\}^{-1}.
\end{equation}
We take this particular expression for $\delta q$ and use it to {\it define} the uncertainty of {\it any} optimal determination of position, including the one considered by Heisenberg in his gedanken experiment. The expression in curly brackets on the right hand side of Eq. (\ref{deltax}) is the Fisher information \cite{F25} associated with translations of $q$. For this reason, $\delta q$ is also known as the the Fisher length \cite{H00}.

As an illustration, consider again a Gaussian distribution. Both the mean and the median are examples of unbiased estimators. But in the case of large samples the variance of the mean is smaller than the variance of the median \cite{Bulmer}, thus the mean would be considered a better estimator. How does the mean compare to other estimators? It is straightforward to show that for a Gaussian distribution, the Cram\'{e}r-Rao lower bound is achieved by the sample mean. Thus it has the lowest possible variance of all unbiased estimators.\footnote{Of course, in the present case, we are not suggesting that the mean of a series of measurements of position be considered as the best estimator, but rather, a single optimal measurement of the type suggested by Heisenberg's microscope.} 

More generally, if an efficient unbiased estimator exists, it will be the maximum likelihood estimator \cite {vdLFvT14}. The expression given by Eq. (\ref{deltax}) is therefore closely connected to maximum likelihood estimation.

\subsection{The uncertainty in momentum}

The experiment considered by Heisenberg is one that measures the location of a particle, thus position plays a fundamental role. Based on the assumption of an optimal measurement, we have defined the uncertainty in position via Eq. (\ref{deltax}). As far as the measurement is concerned, there is an obvious asymmetry between position and momentum, as Hilgevoord and Uffink \cite{HU90} have emphasized,
\begin{quote}
Note that only \emph{one} measurement is actually performed: the determination of the photon's position. From the result of this measurement a prediction can be made about the outcome of \emph{subsequent} measurements of the photon's (or electron's) momentum. No simultaneous measurements are involved! Neither are joint probabilities, nor is the projection postulate.
\end{quote}
Thus the uncertainty in momentum must be handled differently than the uncertainty in position. As there is no measurement of momentum, the uncertainty in momentum cannot be estimated experimentally. Heisenberg writes (using our notation) ``Let [$\delta p$] be the precision with which the value [$p$] is determinable; that is, here, the discontinuous change of $p$ in the Compton effect.'' But this is not sufficient for our purposes. We consider a definition of momentum uncertainty which is both more general and more precise. We will proceed in two steps. We will first consider the case of a classical particle, for which there is a natural and straightforward definition of momentum uncertainty. Afterward, we will look for an appropriate definition of momentum uncertainty that is valid for the quantum case.

\subsubsection{Momentum uncertainty for the classical case}

In the previous section we defined the uncertainty in position $q$ by means of a probability density defined over the configuration space of the particle (i.e., Euclidean three-dimensional space). If the particle is in motion, the probability density associated with the location of the particle will also be in motion. Since momentum is associated with velocity, it is clear that we need to consider the {\it dynamics} of this probability density. There is one fundamental requirement: the probability density $P=P(q,t)$ introduced in the previous section must satisfy a continuity equation, otherwise probability is not conserved. Such an equation is of the form
\begin{equation}\label{contEq}
\dot{P} + \frac{\partial (P v)}{\partial q} = 0
\end{equation}
where $v=v(q,t)$ is the \emph{velocity field} associated with the motion of $P$.

To specify the motion of $P$, we need an equation for the velocity field $v$. We consider first a classical particle, and later we will look at the modifications that are needed for a quantum particle.
In the case of a classical particle, one may define $v$ with the help of the Hamilton-Jacobi formalism, in which case
\begin{equation}\label{velHJ}
v = \frac{1}{m}\frac{\partial S}{\partial q}
\end{equation}
where $S=S(q,t)$ satisfies the the Hamilton-Jacobi equation
\begin{equation}\label{HJEq}
\dot{S} + \frac{1}{2m}\left( \frac{\partial S}{\partial q}  \right)^2 + V = 0,
\end{equation}
with a potential term $V$. It is straightforward to express the average energy $<E>$ associated with the ensemble of classical particles; i.e.,
\begin{equation}\label{EnergyECS}
<E> = \int dq \, P\left[\frac{1}{2m}\left( \frac{\partial S}{\partial q}  \right)^2 + V \right].
\end{equation}
As an aside, notice that it is not strictly necessary to postulate Eq. (\ref{velHJ}) because $v$ follows from the equations of motion for $P$ and $S$ whenever $P$ is highly localized (i.e., a delta function). Thus $v$ is operationally well defined
\cite{HR05,HR16}.
If $P$ is not localized then
$P\partial S/\partial q$ can be thought of as momentum density associated with $P$.

It will be useful for the analysis of the next section to restrict the formulation to the specific case of interest. Since the gedanken experiment involves a free particle immediately before and after the collision, we are interested in the case in which the potential term is set to zero. From Eq. (\ref{EnergyECS}), the average energy for the case $V=0$ is given by
\begin{equation}\label{HamiltonianECS}
H_{C}=\int dq \, P\left[\frac{1}{2m}\left( \frac{\partial S}{\partial q}  \right)^2 \right].
\end{equation}
We will see in the next Section that $H_{C}$ plays the role of a {\it classical Hamiltonian}; i.e., the Hamiltonian that determines the equation of motion for the ensemble of free classical particles.

Now we can \emph{define} the uncertainty in the momentum $p_{C}$ of a classical particle in the usual way via the variance,
\begin{equation}\label{varPC}
(\delta p_C)^2 = var(p_{C}) = \left\{ \int dq \, P \,  \left( \frac{\partial S}{\partial q}  \right)^2 \right\} - \left\{ \int dq \, P \, \left( \frac{\partial S}{\partial q} \right) \right\}^2 ,
\end{equation}
where we use Eq. (\ref{velHJ}) to express the momentum in terms of $\frac{\partial S}{\partial q}$.  Notice that $var(p_{C})$ and $H_C$ are related by the equation
\begin{equation}\label{varPandHC}
var(p_{C}) = <(p_C)^2> - <p_C>^2 = 2mH_C - <p_C>^2.
\end{equation}
This observation will be useful in the next section.

What changes do we have to make to this formulae when we have a {\it quantum} particle? Clearly the expression for
$(\delta p_C)^2$, Eq. (\ref{varPC}), can no longer be valid because it is inconsistent with the Heisenberg uncertainty relation. While Eq. (\ref{contEq}), the continuity equation, must be satisfied to ensure conservation of probability, the classical Hamilton-Jacobi equation, Eq. (\ref{HJEq}), certainly cannot be applied to quantum phenomena. Both of these inadequacies can be corrected by adding a quantum term to the classical Hamiltonian, a task to which we now turn.

\subsubsection{Momentum uncertainty for the quantum case}

The previous steps did not require anything particularly out of the ordinary. The uncertainty in position $\delta q$ was defined using tools of statistics developed in the 1920s for precisely such types of measurement. The uncertainty in momentum $\delta p_{C}$ of the classical particle was defined by means of Hamilton-Jacobi theory and some basic concepts from probability theory. The next step of jumping to the Schr\"{o}dinger equation requires more creativity.

Heisenberg's analysis leads to the semiquantitative expression $\delta q \, \delta p \sim h$, which should be valid for the case of an \emph{optimal} measurement carried out on a quantum particle  if the experiments are to be consistent with the claim that this relation represents a {\it fundamental limit}. We rewrite this as the equality
\begin{equation} \label{heq}
\delta q \, \delta p_{H} = \frac{h}{\eta},
\end{equation}
where $\eta > 0$ is a dimensionless constant which Heisenberg did not bother to determine and the subscript $H$ in  $\delta p_{H}$ is there to remind us that this is the uncertainty in momentum that appears in Heisenberg's indeterminacy relation,  Eq. (1).  Eq. (\ref{heq}) can be written in the equivalent form
\begin{equation} \label{vph}
(\delta p_H)^2 = \frac{h^2}{\eta^2}\frac{1}{(\delta q)^2}=\frac{h^2}{\eta^2} \int dq \, P  \left(\frac{1}{P} \frac{\partial P}{\partial q} \right)^2 ,
\end{equation}
where we used Eq. (\ref{deltax}) in the last equality. Furthermore, since Heisenberg considers a ``best case'' scenario, the product $\delta q \delta p$ will typically be greater than $\frac{h}{\eta}$.
So we replace Heisenberg's relation by the inequality
\begin{equation} \label{hineq}
\delta q \, \delta p_Q \geq \frac{h}{\eta},
\end{equation}
which presumably is valid for any measurements performed on a quantum particle.  The symbol $\delta p_Q$ represents the variance in subsequent measurements of the momentum of the quantum particle.  This may seem somewhat removed of the $\delta p$ in Heisenberg's indeterminacy relation, which represented a disturbance of the momentum caused by the $\gamma$-ray microscope.  However, one can certainly claim that this disturbance is equivalent to the uncertainty in any subsequent measurement of the particle's momentum.  In this sense, it is reasonable to consider the use of the $\gamma$-ray microscope to be a procedure of state preparation going forward.  Heisenberg seemed to make no distinction between these two interpretations of $\delta p$.

We would like to find a reasonable expression for $\delta p_Q$ which involves $\delta p_C$ and $\delta p_H$, the only two other uncertainties that we have at our disposal. In the optimal quantum limit $\delta p_Q \rightarrow \delta p_H$, while in the classical limit $\delta p_Q \rightarrow \delta p_C$.
The simplest way to agree with these two limits is to {\it define} $(\delta p_Q)^2$ as the {\it sum} of $(\delta p_C)^2$ and $(\delta p_H)^2$; i.e., $\delta p_Q$ is obtained by summing $\delta  p_C$ and $\delta p_H$ in quadrature. While this step has no rigorous justification, it is consistent with the standard propagation of uncorrelated errors, a statistical procedure for data analysis that  dates back to Gauss. With this assumption,
\begin{eqnarray}\label{varPQ}
(\delta p_Q)^2 &=& ( \delta p_C)^2 + (\delta  p_H)^2 \nonumber\\
&=&   \int dq \, P \,  \left( \frac{\partial S}{\partial q}  \right)^2  - \left[ \int dq \, P \, \left( \frac{\partial S}{\partial q} \right) \right]^2  +  \frac{h^2}{\eta^2} \int dq \, P \left(\frac{1}{P}  \frac{\partial P}{\partial q} \right)^2 \nonumber\\
&=& \left\{ \int dq \, P \,  \left[\left( \frac{\partial S}{\partial q} \right)^2 + \frac{h^2}{\eta^2} \left( \frac{1}{P} \frac{\partial P}{\partial q} \right)^2 \right] \right\} - \left\{ \int dq \, P \, \left( \frac{\partial S}{\partial q} \right) \right\}^2\nonumber\\.
\end{eqnarray}
We now set $var(p_Q)=(\delta p_Q)^2$. This is the  expression for $var(p_Q)$ that ought to replace the classical variance $var(p_C)$ for the case of a quantum particle. We now look at the consequences of defining $var(p_Q)$ in terms of the $(\delta p_Q)^2$ of Eq. (\ref{varPQ}).

\subsection{From the Heisenberg uncertainty relation to the Schr\"{o}dinger equation}\label{IIIC}

From Eq. (\ref{varPandHC}), $var(p_C)=2mH_C-<p_C>^2$. If we assume a similar functional form for $var(p_Q)$ we are led to a quantum Hamiltonian equal to the first term in square brackets of the last equality of Eq. (\ref{varPQ}) divided by $2m$,
\begin{equation}\label{HQ}
H_Q =   \int dq \, P \,  \left[ \frac{1}{2m}\left( \frac{\partial S}{\partial q} \right)^2 + \frac{h^2}{2m\eta^2} \frac{1}{P^2} \left( \frac{\partial P}{\partial q} \right)^2 \right].
\end{equation}
This expression for $H_Q$ replaces the classical Hamiltonian $H_C$ for the case of a quantum particle.  Eq. (\ref{varPandHC}) results from identifying the classical Hamiltonian with the kinetic energy of a free particle.  If one takes into account the increased quantum dispersion implied by the uncertainty relation, it is clear that the effective kinetic energy of a quantum particle is the classical expression plus the dispersion of Eq. (\ref{vph}) divided by $2m$.  Equating this effective kinetic energy with the quantum Hamiltonian leads to  and provides further justification for Eq. (\ref{HQ}).

Now comes a crucial observation: it is straightforward to show that the two equations that describe the ensemble for a classical particle, Eq. (\ref{contEq}) and Eq. (\ref{HJEq}), can be derived using a Hamiltonian formalism if we take $P$ and $S$ as canonically conjugate field variables and define rate equations in the usual way,
\begin{eqnarray}
\dot{P} &=& \{P,H_{C}\} = -\frac{\partial (P v)}{\partial q},\\
\dot{S} &=& \{S,H_{C}\} = -\frac{1}{2m}\left( \frac{\partial S}{\partial q}  \right)^2,
\end{eqnarray}
where $\{F,G\}$ denotes the Poisson bracket of $F$ and $G$. We will now assume that the equations that describe the ensemble for a \emph{quantum} particle will follow in an analogous way, but with $H_C$ replaced by $H_Q$.

The equations that follow from the Hamiltonian $H_Q$ are the continuity equation, Eq. (\ref{contEq}), and the modified Hamilton-Jacobi equation,
\begin{equation}\label{MHJEq}
\dot{S} + \frac{1}{2m}\left( \frac{\partial S}{\partial q}  \right)^2 + \frac{4 h^2}{2m \eta^2}\frac{1}{\sqrt{P}}\frac{\partial^2 \sqrt{P}}{\partial q^2}= 0.
\end{equation}
This equation replaces the classical Hamilton-Jacobi equation for the case of a quantum particle.  As expected, when $\eta \rightarrow \infty$ Eq. (22) becomes the classical Hamilton-Jacobi equation.

At this point, $\eta$, is still a free parameter, one that Heisenberg left undefined. In principle, it could be determined experimentally. One might attempt to determine the value of $\eta$ by a more careful analysis of Heisenberg's microscope experiment or by considering another more optimal arrangement.  However, this would certainly go beyond Heisenberg's motive; he intentionally left the symbol `$\sim$' in his relation, $\delta q \delta p \sim h$.  Alternatively, one could leave $\eta$ as an unknown factor and then compare the results of subsequent calculations with known phenomena such as atomic spectra.  This would inevitably result in determining $\eta$ to be $4 \pi$. If we set $\eta=4 \pi$, Eq. (\ref{contEq}) and Eq. (\ref{MHJEq}) are precisely the Schr\"{o}dinger equation written in the formulation that  Madelung introduced in 1926 \cite{M26}. The complex Madelung transformation, $\psi=\sqrt{P}e^{iS/\hbar}$, maps these two real equations to the free particle Schr\"{o}dinger equation in its usual form,
\begin{equation}\label{SEq}
i \hbar \dot{\psi} = - \frac{\hbar^2}{2m}\frac{\partial^2 \psi}{\partial q^2}.
\end{equation}

In this way, we have outlined a path from the Heisenberg uncertainty relation to the Schr\"{o}dinger equation for a free particle and, perhaps, fulfilled Heisenberg's  wish ``...to be able to derive, if possible, the quantitative laws of quantum mechanics directly from the physical foundations - that is, essentially, from $\delta q \delta p \sim h$.''

A claim to have {\it derived} the Schr\"{o}dinger equation would be a bit of an overstatement considering the assumptions made above, especially the momentum variance of Eq. \ref{varPQ}.  The choice for the quantum Hamiltonian in Eq. \ref{HQ} seems less worrisome given that the freedom to choose appropriate Hamiltonians is often exercised.  Even so, we find the self-consistency of our approach convincing.  For example, the substitution of the Madelung expression for $\psi$ in the standard quantum mechanical expression for momentum variance,
$\int dq \, \psi^*(\frac{\hbar}{i}\frac{\partial}{\partial q})^2 \psi - \left[ \int dq \,\psi^* \frac{\hbar}{i}\frac{\partial}{\partial q} \psi \right] ^2$, yields precisely the variance in Eq. \ref{varPQ}.  Again using the Madelung transformation but going in the other direction, the Hamilton-Jacobi expression for the average momentum, $\langle p\rangle =\int dq \, P\, \frac{\partial S}{\partial q}$  implies the quantum mechanical expression $\langle p\rangle =\int dq \, \psi^*\frac{\hbar}{i}\frac{\partial}{\partial q} \psi$.   Another example is the quantum mechanical minimum uncertainty (Gaussian) wave packet.  Again from the Madelung transformation, it's straightforward to show that the equivalent Hamilton-Jacobi probability density for such a packet is $\frac{e^{-q^2/2\sigma^2}}{\sqrt{\pi \sigma^2}}$, in which case the variance of $q$ is $\delta q^2=\frac{\sigma^2}{2}$, the lower bound of the Cram\'{e}r-Rao inequality of Eq, (\ref{deltax}), while the momentum variance of Eq. (\ref{varPQ}) for such a packet is $\delta p^2=\frac{\hbar^2}{2\sigma^2}$, resulting in $\delta q \delta p = \frac{\hbar}{2}$, our assumed minimum Heisenberg uncertainty.

\section{Discussion}

The analysis of Section 3 maps out a direct path from Heisenberg's indeterminacy relation to the formalism of quantum mechanics as embodied in the Schr\"{o}dinger equation.  Because the analysis was framed in terms of probabilities, the statistical interpretation is automatic and there is no need to introduce the Born Rule for interpreting the wave function $\Psi$.  As such, it provides support for Heisenberg's conclusion (as stated in the abstract of his 1927 paper), i.e., ``This indeterminacy is the real basis for the occurrence of statistical relations in quantum mechanics.''
Our analysis also appears to fulfill, in part,  Heisenberg's desire to derive the laws of quantum mechanics from his indeterminacy relation, $\delta q \delta p \sim h$. The endpoint of our analysis, the Schr\"{o}dinger equation for a free particle, automatically includes the wave behavior of matter with no additional assumptions.

It is instructive to review  how this happens. Heisenberg shows us that there are inevitably uncertainties $\delta q$ and $\delta p$ associated with the position and momentum of a particle. Once the uncertainty in position is accepted as a fundamental aspect of the theory, it is natural to describe the position of the particle by a probability density $P$ that is mathematically a field over configuration space.
Since the particle moves, $P$ will have to change with time, and since there is no reason why it should move in a rigid fashion, there has to be a second field, a velocity vector field $\vec{v}$, which describes how the probability $P$ changes in time. As we have shown in section 3, the classical limit (in the form of the Hamilton-Jacobi formalism) suggests that the velocity field can be derived from a single  scalar  field $S$ according to $m \vec{v}=\nabla S$. Thus the motion of the particle is conveniently described in terms of two fields, $P$ and $S$, {\it as a consequence of the indeterminacy postulated by Heisenberg}. The derivation of the Schr\"{o}dinger equation from this insight is of course not a trivial exercise, as is apparent from the analysis of section 3, but it can be carried out.
Heisenberg introduced his $\gamma$-ray microscope to investigate the very meanings of {\it position} and {\it momentum} of a particle.  He maintained that these concepts only derive meaning by virtue of experiments with which they are to be measured.  The result of his deliberations was the indeterminacy relation, $\delta q \delta p \sim h$.  It should be emphasized that this relation is quite distinct from the Kennard/Robertson/Schr\"{o}dinger uncertainty principle, $\Delta q \Delta p \geq \hbar/2$.  The latter is derived directly from the formalism of quantum mechanics and is a statement about the properties of quantum states.  The quantum states themselves can be associated with hypothetical measurements only via the Born rule and interpretative statements relating quantum operators and the outcome of these measurements, the descriptions of which are usually vague and lie outside the formalism of quantum mechanics.  Heisenberg's indeterminacy relation, on the other hand, is a statement about the very meanings of $q$ and $p$ as derived from our ability to determine them using an ideal measuring apparatus.  It is in this sense that $\delta q \delta p \sim h$ is an empirical principle of nature. It is remarkable that the definition of $\delta q$ that is needed to go from the Heisenberg uncertainty relation to the Schr\"{o}dinger equation, Eq. (\ref{deltax}), is associated with maximum likelihood estimation. This suggests a non-trivial connection between the Schr\"{o}dinger equation and classical statistics which should be explored further.

Since the early days of quantum mechanics there have been many discussions as to how the theory should be interpreted.  It seems likely that these questions arise in part as a consequence of adopting the formalism of quantum mechanics, e.g., the Schr\"{o}dinger equation, as primal.  One is then left with the problem of identifying the theoretical constructs, the wave functions or quantum state vectors, with aspects of reality.  On the other hand, if one considers Heisenberg's empirical indeterminacy relations as primal, such questions do not immediately arise.  Then the formalism of quantum mechanics and concomitant wave functions become simply tools for making predictions.

In his 1927 paper, Heisenberg maintained that the statistical nature of quantum mechanics arises from the indeterminacy relations. It should be noted that our analysis in Section 3 also treated classical physics as probabilistic.  That statistics is important to classical physics is not particularly novel; however, classical indeterminacy is usually relegated to the behavior of large numbers of particles (statistical mechanics) or simply labeled as experimental noise and then summarily dismissed as a fundamental aspect of nature.  This doesn't have to be the case, as indicated by the analysis of Section 3.  So if it's not its statistical nature that distinguishes quantum from classical mechanics, what does? It is generally accepted that there is no classical analog of the wave nature of particles nor of the concomitant quantum interference. We have shown that the Schr\"{o}dinger equation, which provides a wave description, can be derived from Heisenberg's indeterminacy relations.  
The implication seems to be that it is primarily the quantal nature of the world that leads to quantum interference and so distinguishes quantum from classical mechanics.

Heisenberg's uncertainty principle, even in its expression as an inequality, is generally acknowledged as describing an underlying property of all quantum mechanical systems.  Even so, there are conflicting claims as to its fundamental importance as well as to its general applicability.  We have here endeavored to argue that the uncertainty relation is of primal importance to the foundations of quantum mechanics, a view that is certainly expressed by Heisenberg in his 1927 seminal paper.

\ack{We thank M. J. W. Hall for valuable discussions.}

\end{document}